\documentclass[12pt,a4paper]{article}
\usepackage[utf8]{inputenc}
\usepackage{amsmath}
\usepackage{amsfonts}
\usepackage{amssymb}
\usepackage{mathtools}
\usepackage{graphicx}
\usepackage{cite}
\usepackage[capitalise]{cleveref}
\graphicspath{ {./figures/} }
\usepackage[left=2cm,right=2cm,top=2cm,bottom=2cm]{geometry}
\crefformat{section}{#2section~#1#3}

\author{Tetiana Obikhod and Ievgenii Petrenko}
\title{\bf{Computer modeling of $t\bar{t}h(H)$ Higgs boson production with MSSM model}}
\date{%
    {\it Institute for Nuclear Research NAS of Ukraine, Kyiv 03028, Ukraine}\\%
    \today
}
\begin{document}

\maketitle

\section{Abstract}

	We studied the properties of the MSSM Higgs bosons, h and H through the decay into b-quarks in associated production with a top-quark pair. There was used the tree-level Higgs sector described by two parameters M$_A$ and tan$\beta$  and found their optimal values according to experimental data of ATLAS detector. Using the restricted parameter space we calculated cross sections of associated $t\bar{t}h(H)$ production at 13 and 14 TeV, the corresponding kinematical cuts, mass distributions and Branching Ratios of h and H decays into $b\bar{b}$ quark pair. 

\section{Introduction}

    Supersymmetry searches are the most attractive in the aspect of searching for new physics beyond the standard model. The search for an extended sector of Higgs bosons is especially urgent, since they are the lightest candidates for supersymmetric particles and information on their production cross sections and decay widths provides additional knowledge about the Yukawa coupling constants. The dependences on these couplings of the cross section and Higgs branching ratios have been studied intensively and it was found, that there are indirect constraints from experimental data on the scalar and 
pseudoscalar H-top couplings k$_t$ and $\tilde{k}$, and these constraints are relatively weak, \cite {1.}  
	
	One of the important channels of such searches is the $t\bar{t}H$ Higgs boson production channel. The search strategy for the $t\bar{t}H$ process has been studied in various Higgs decay modes: $b\bar{b}$, \cite{2.}, $\tau\bar{\tau}$, \cite{3.} and WW$^*$, \cite{4.}.  Furthermore, from experimental point of view the $H\rightarrow b\bar{b}$ decay mode is more prefferable due to the possibility of the reconstruction of the Higgs boson kinematics, which allows to extract the information about the top–Higgs interaction. 
    As the decay of Higgs boson into two b-quarks ($H\rightarrow b\bar{b}$) is the most probable, \cite{5.} our paper is devoted to the consideration of this decay channel. Higgs boson decay into b-quarks in associated production with a top-quark pair is connected with
testing the predictions of the Standard Model (SM) and
very sensitive to effects of physics beyond the SM (BSM). So,
we used Minimal Supersymmetric Standard Model (MSSM) as base theory for futher calculations. Our purpose was to calculate production cross sections $\sigma(t\bar{t}h),\ \sigma(t\bar{t}H)$ at the centre-of-mass energy of $\sqrt{s}$ = 13 TeV and to compare obtained data with experimental data. We also found p$_T$ and rapidity distributions, parameter space and mass distributions, which corresponds to the best fit with experimental data.

\section{Search channels and parameter cuts of Higgs boson production}
	ATLAS \cite{6.} and CMS \cite{7.} have searched for the process of  Higgs boson production ($t\bar{t}H(b\bar{b})$)  intensively using the 8 TeV data set. Later new data collected in proton–proton collisions at the LHC between 2015 and 2018 at a centre-of-mass energy of $\sqrt{s}$ = 13 TeV were analysed, corresponding to an integrated luminosity of 139 fb$^{-1}$, \cite{8.}. The measured signal strength, defined as the ratio of the measured signal yield to that predicted by the SM, 
\[ \mu = 0.35\pm0.20 (stat.)^{+0.30}_{-0.28}(syst.)=0.35^{+0.36}_{-0.34} \ ,\]
corresponds to an observed (expected) significance of 1.0 (2.7) standard deviations. The measured 95$\%$ confidence level (CL) cross-section upper limits in each bin for simplified template cross-sections (STXS) formalism are shown in Fig.1
\begin{center}
\includegraphics[width=0.4\textwidth]{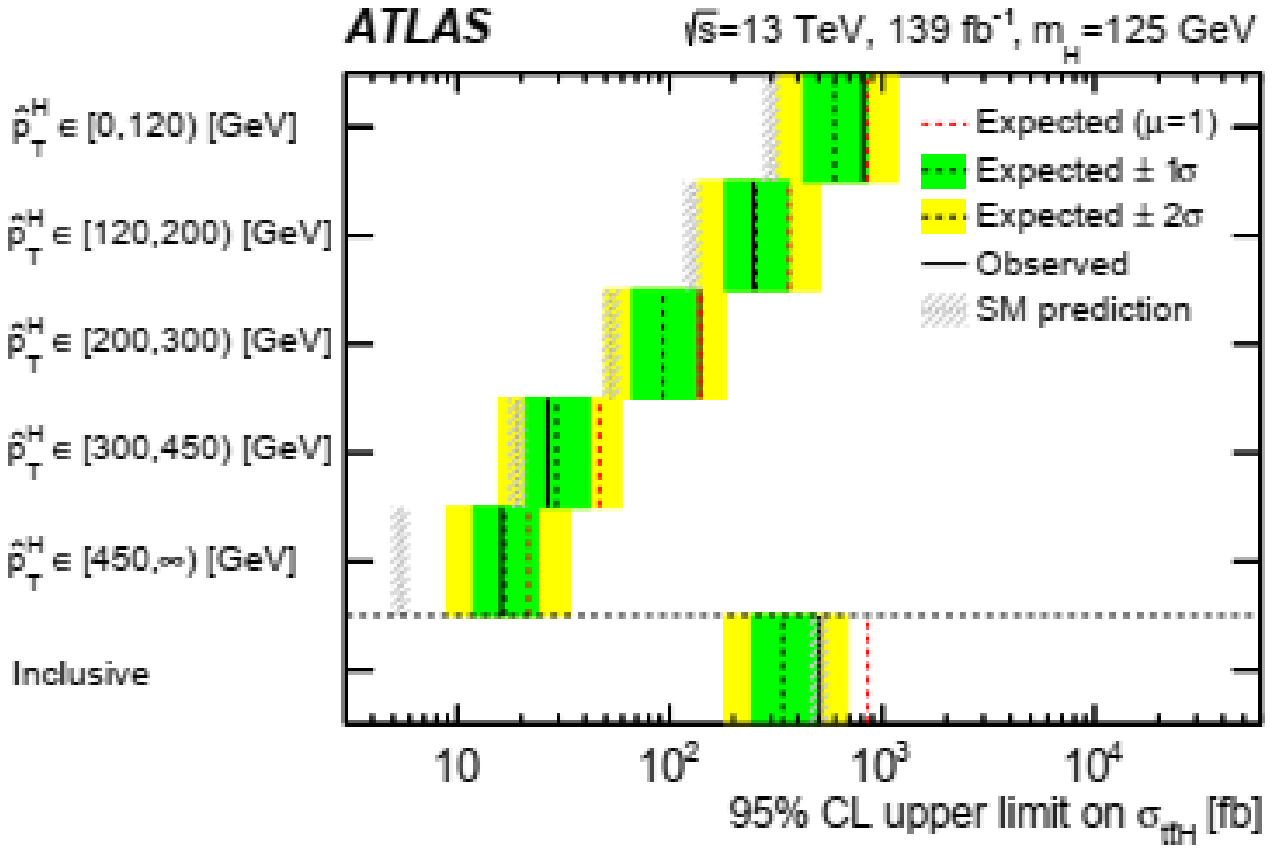}\\
\emph{\textbf{Fig.1}} {\emph{The measured 95$\%$ CL cross-section upper limits with the theoretical uncertainty 
connected with signal scale and PDF uncertainties.}}
\end{center}

    The Higgs sector of the Minimal Supersymmetric extension of the SM \cite{9.} consists of five physical Higgs bosons, two neutral CP-even bosons, h, H, one neutral CP-odd boson, A, and a charged Higgs pair, H$^{\pm}$. 
As there is the experimental deviation from the SM, we used BSM model - MSSM and studied the properties of two Higgs bosons: the lightest Higgs boson, h and CP-even Higgs boson, H. Our analysis of  Higgs boson production in association with a pair of top quarks and decaying into a pair of b-quarks ($t\bar{t}H(b\bar{b})$) is presented by Feynman diagrams in Fig.2.
\begin{center}
\includegraphics[width=0.5\textwidth]{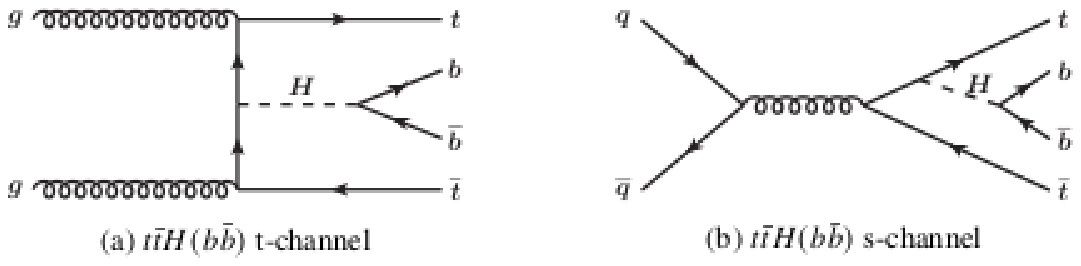}\\
\emph{\textbf{Fig.2}} {\emph{Representative tree-level Feynman diagrams for the production of a Higgs boson in association with a
top-quark pair ($t\bar{t}H$) in (a) the t-channel and (b) the s-channel and the subsequent decay of the Higgs boson into $b\bar{b}$, from \cite{8.}.
}}
\end{center}
The tree-level Higgs sector, can be described by two parameters, the mass of the CP-odd Higgs boson, M$_A$, and the ratio of the two vacuum expectation values of the two Higgs doublets, tan$\beta = v_2/v_1$. So, our purpose was to chooose the optimal value of the correspomding parameters M$_A$ and tan$\beta$. We calculated Branching ratio (BR) of h and H using FeynHiggs program \cite{10.} as the function of M$_A$ at 13 TeV (Fig. 3) as well as production cross section as the function of tan$\beta$ at M$_A$=200 GeV (Fig.4).
\begin{center}
\includegraphics[width=0.5\textwidth]{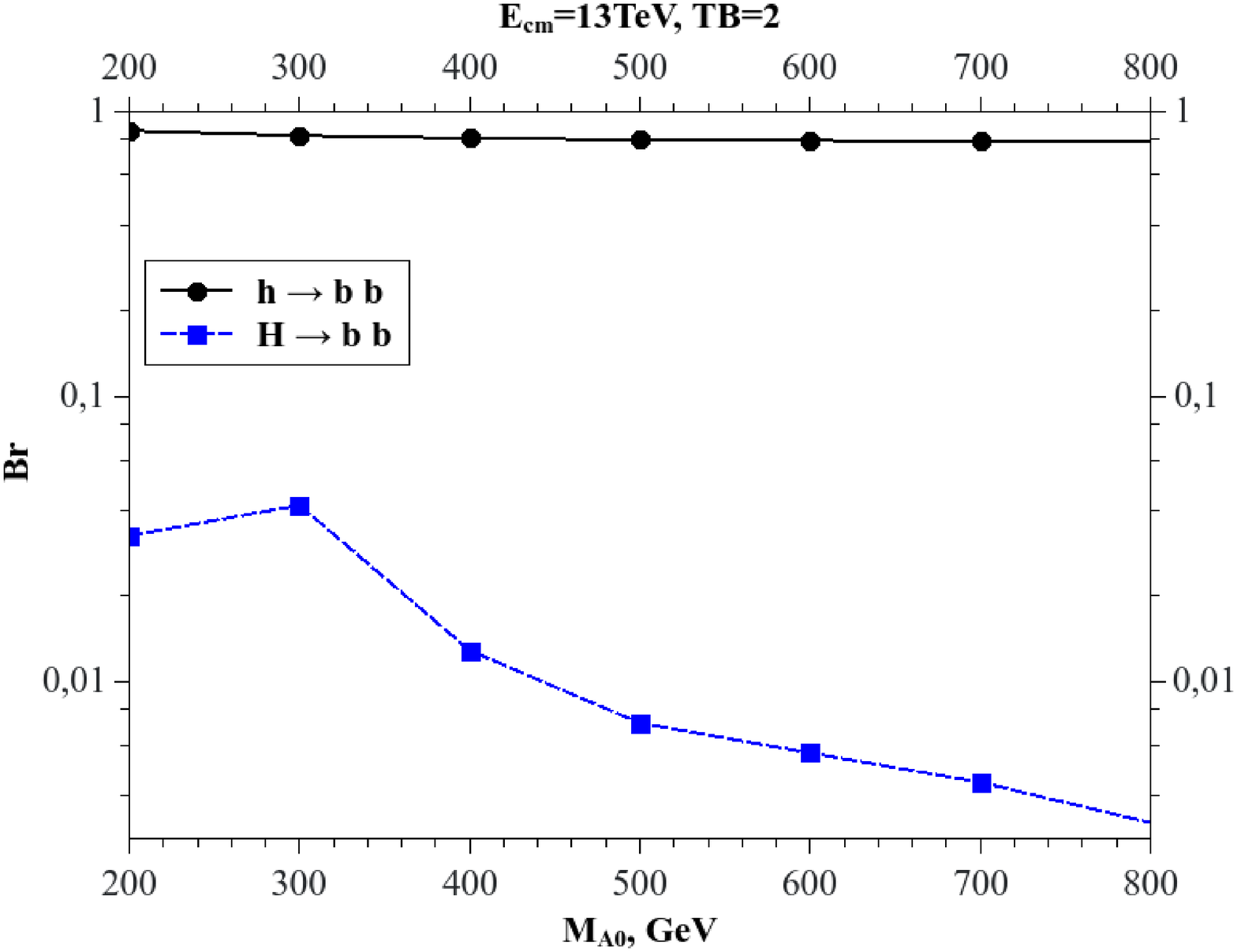}\\
\emph{\textbf{Fig.3}} {\emph{Branching ratio (BR) of h and H decays using FeynHiggs program as the function of M$_A$ at 13 TeV.}}
\end{center}
\begin{center}
\includegraphics[width=0.5\textwidth]{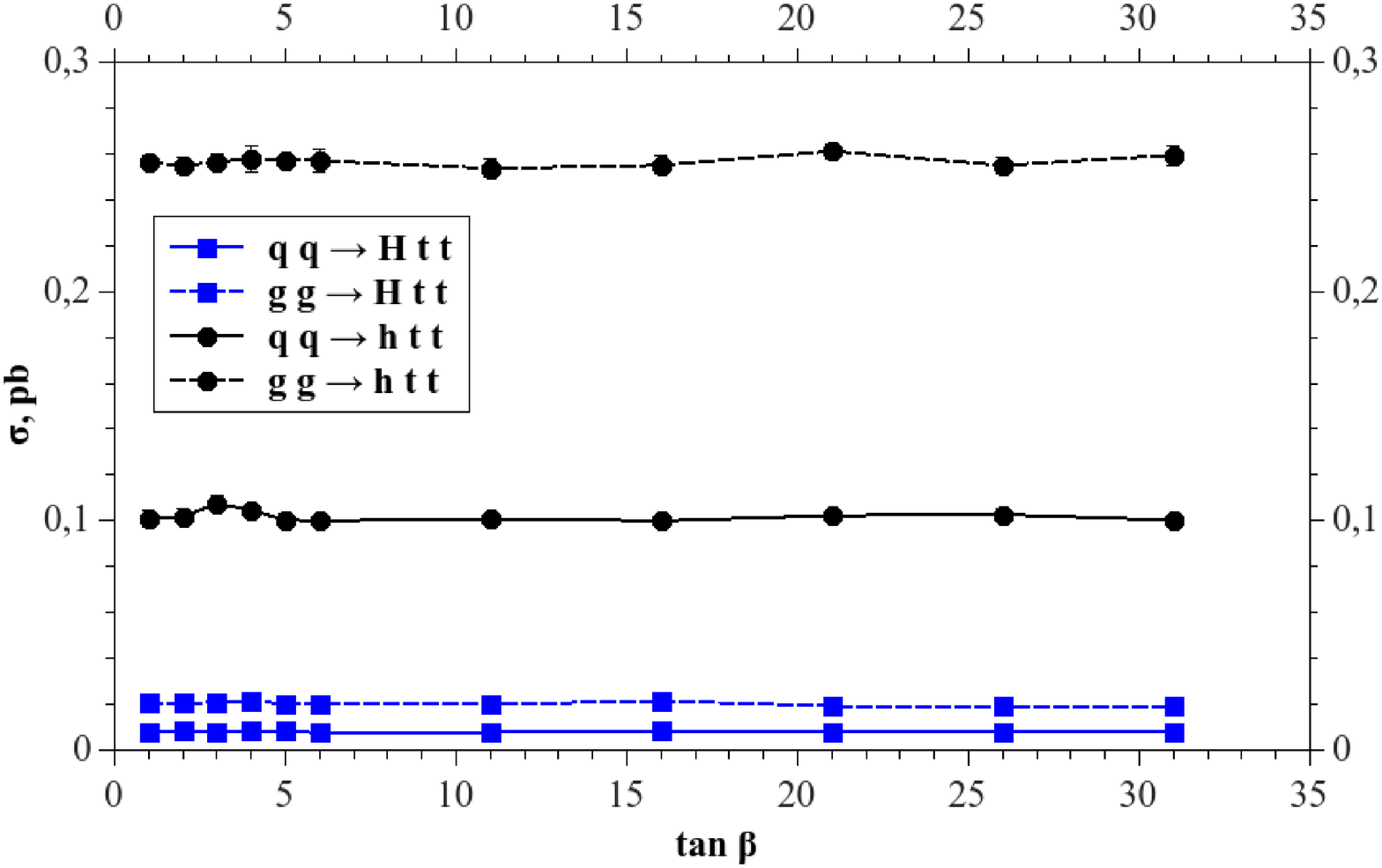}\\
\emph{\textbf{Fig.4}} {\emph{Production cross section of Higgs bosons h and H for two search channels as the function of tan$\beta$ at M$_A$=200 GeV at 13 TeV.}}
\end{center}
    
From the obtained data we came to the conclusion about the optimum parameters of M$_A$=200 GeV and tan$\beta$ =2.

\section{Results of calculations}
    As gluon-gluon and quark-antiquark processes are most preferable for the Higgs boson production we have considered the following search channels using Pythia program \cite{11.}, presented in Table 1 and Table 2

\begin{center}
{\it\normalsize Table 1. Search channels and production cross sections of h and H bosons at the centre-of-mass energy of $\sqrt{s}$ = 13 TeV at M$_A$ = 200 GeV and tan$\beta$ =2}\\
\vspace*{3mm}
\begin{tabular}{|c|c|} \hline 
Search channels& Production cross sections (pb) \\ 
& (with stat. err.) \\ \hline \hline
gg $\rightarrow$  $ht\bar{t}$ &   2.609e-01 +/- 5.084e-03   \\ \hline
$q\bar{q}$ $\rightarrow$  $ht\bar{t}$   &   1.034e-01 +/- 1.287e-03
   \\ \hline
gg $\rightarrow$  $Ht\bar{t}$     &   2.017e-02 +/- 5.618e-04   \\ \hline
$q\bar{q}$ $\rightarrow$ $Ht\bar{t}$   &    7.524e-03 +/- 3.862e-04
   \\ \hline
     gg $\rightarrow$  $Ht\bar{t}$ (SM)    &    2.530e-1 +/- 2.704e-3  \\ \hline
$q\bar{q}$ $\rightarrow$  $Ht\bar{t}$ (SM)      &    1.041e-1 +/-  2.572e-3
   \\ \hline
\end{tabular}
\end{center}

\begin{center}
{\it\normalsize Table 2. Search channels and production cross sections of h and H bosons at the centre-of-mass energy of $\sqrt{s}$ = 14 TeV at M$_A$ = 200 GeV and tan$\beta$ =2}\\
\vspace*{3mm}
\begin{tabular}{|c|c|} \hline 
Search channels& Production cross sections (pb) \\ 
& (with stat. err.)\\ \hline \hline
gg $\rightarrow$ $ht\bar{t}$      &  3.134e-01 +/- 4.095e-04   \\ \hline
$q\bar{q}$ $\rightarrow$  $ht\bar{t}$   &  1.171e-01 +/- 1.660e-04  \\ \hline
gg $\rightarrow$ $Ht\bar{t}$      &  2.514e-02 +/- 6.099e-05   \\ \hline
$q\bar{q}$ $\rightarrow$  $Ht\bar{t}$ &  9.221e-03 +/- 5.913e-05 \\ \hline
     gg $\rightarrow$ $Ht\bar{t}$ (SM)          &  2.530e-1 +/- 2.704e-3   \\ \hline
$q\bar{q}$ $\rightarrow$  $Ht\bar{t}$ (SM)      &  1.041e-1 +/- 2.572e-3  \\ \hline
\end{tabular}
\end{center}

The  kinematical cuts on h boson corresponding to the production cross section are presented in Fig. 5.

\begin{center}
\includegraphics[width=0.4\textwidth]{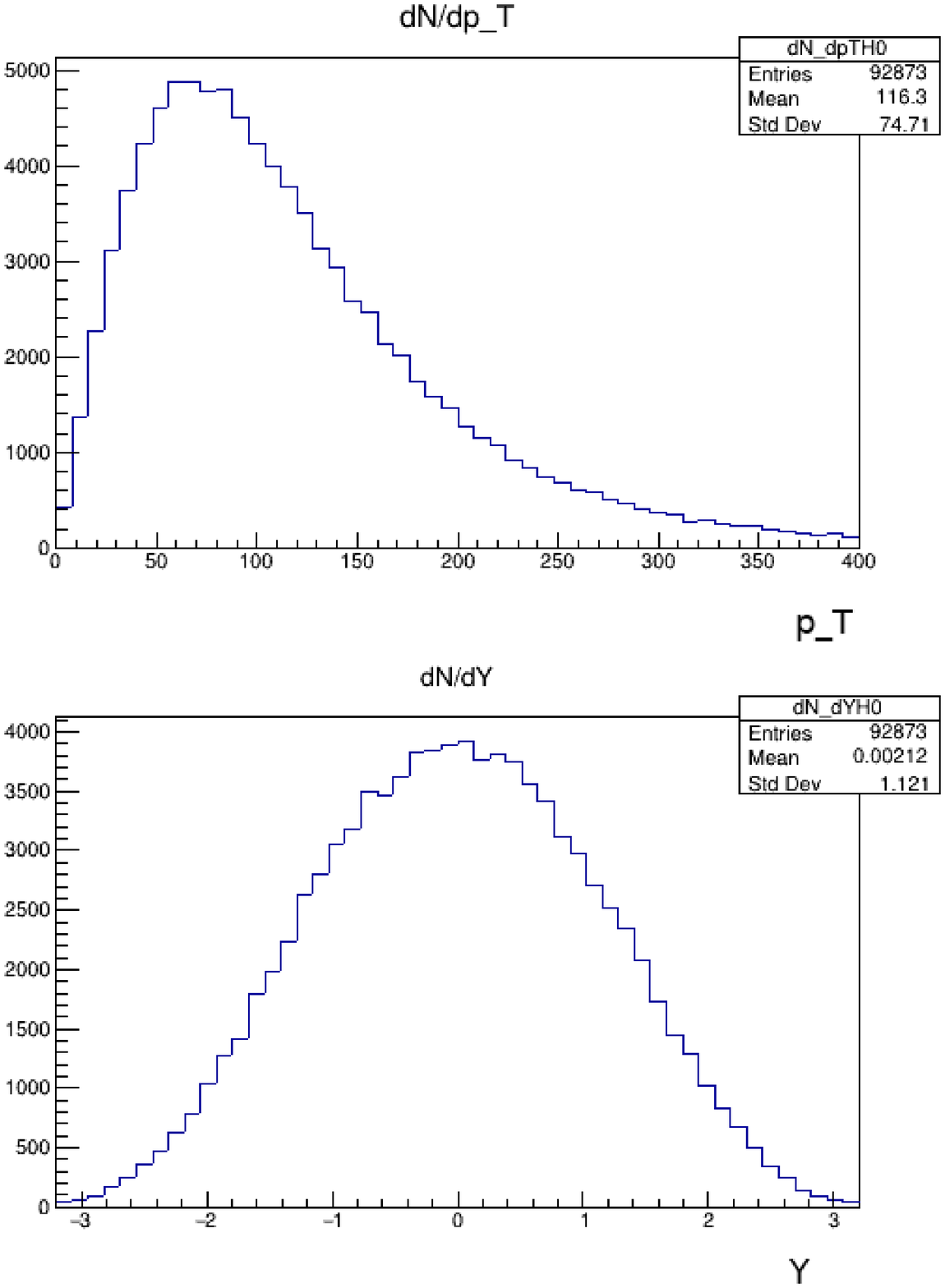}\\
\emph{\textbf{Fig.5}} {\emph{Transverse momentum (up) and rapidity distributions (down) of h boson at the energy of 13 TeV.}}
\end{center}
The  kinematical cuts on H boson corresponding to the production cross section are presented in Fig. 6.
\begin{center}
\includegraphics[width=0.4\textwidth]{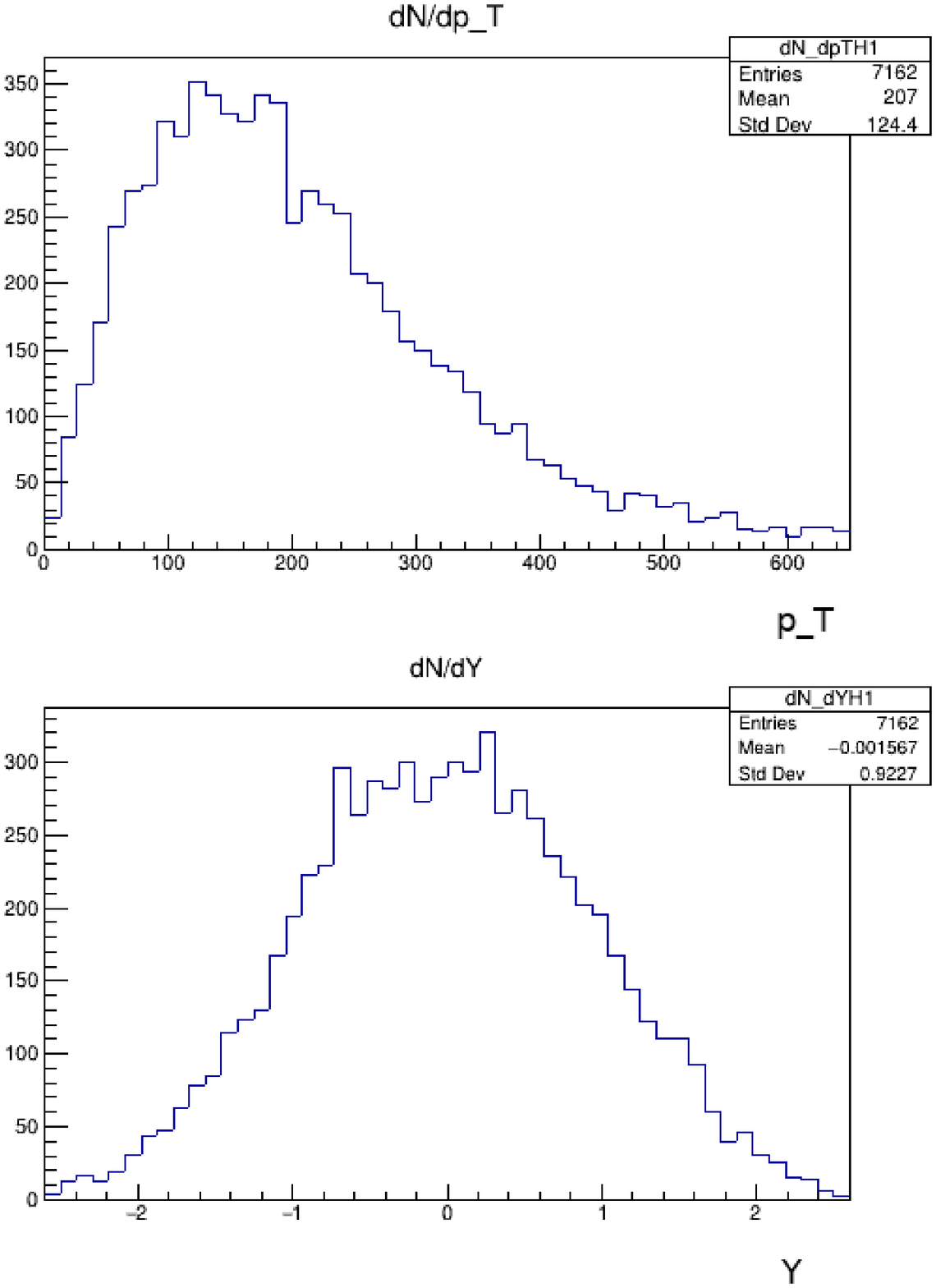}\\
\emph{\textbf{Fig.6}} {\emph{Transverse momentum (up) and rapidity distributions (down) of H boson at the energy of 13 TeV.}}
\end{center}
From the obtained data we came to the conclusion about the most suitable range of transverse momentum variation of h boson (50; 150) GeV and (100; 300) GeV of H boson. As for rapidity distributions of both Higgs bosons, their maximum region is  (-1,1), which signals about the angle range along the longitudinal (beam) direction ~ 45-90$^{\circ}$ of the Higgs bosons. 
	
    As for the mass detemination of both Higgs bosons, we received the mass distributions, presented in Fig. 7
\begin{center}
\includegraphics[width=0.4\textwidth]{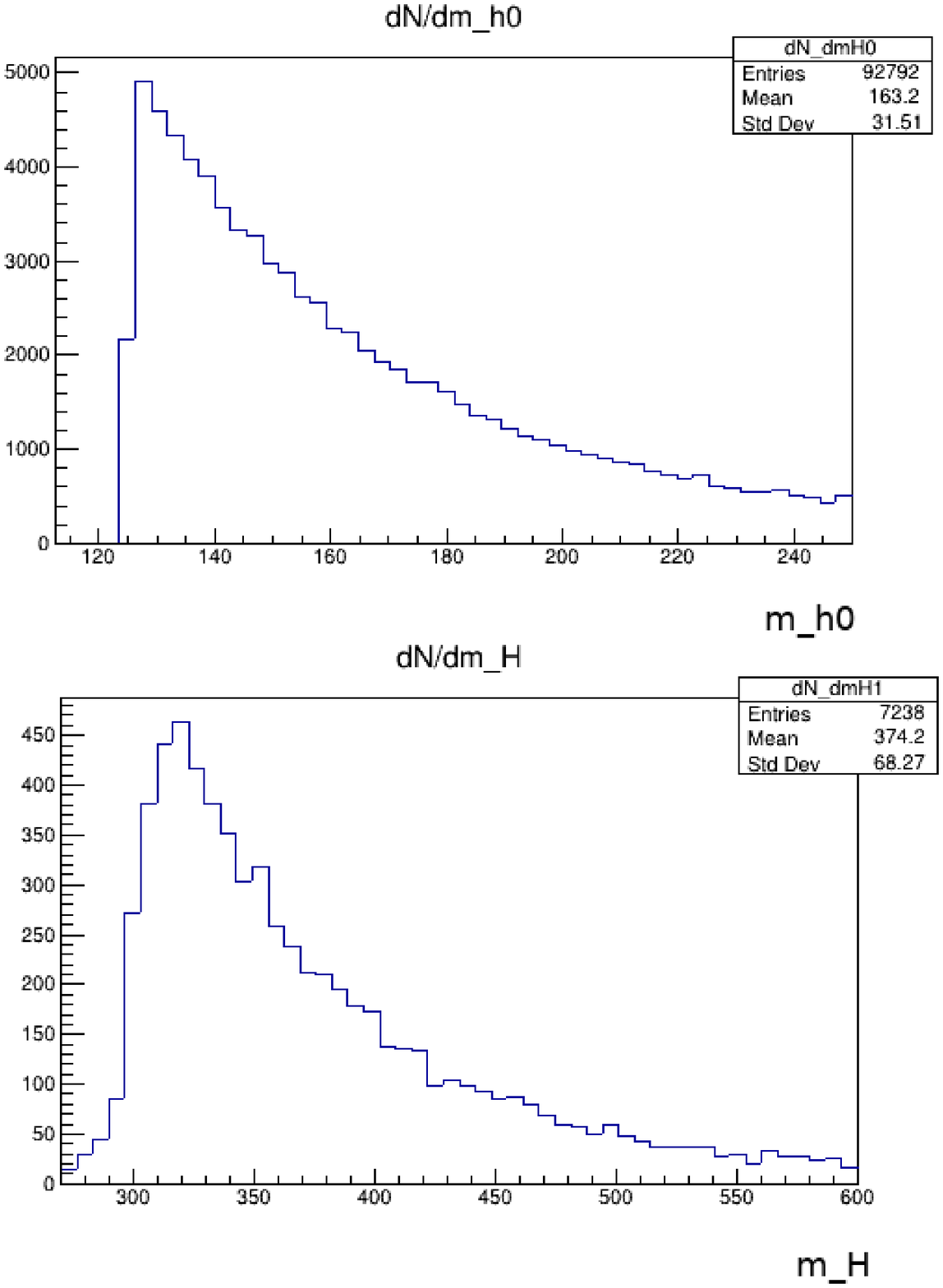}\\
\emph{\textbf{Fig.7}} {\emph{Mass distributions of h (up) and H (down) Higgs bosons obtained at 13 TeV.}}
\end{center}
From Fig.7 we see, that the mass of h boson is about 126 GeV, which coincides with the mass of the SM Higgs boson. As for the H boson, it's mass is approximately 330 GeV. We also calculated kinematical and mass distributions of both Higgs bosons at 14 TeV and didn’t find a significant difference with previous calculations at 13 TeV.
 \section{Conclusions}

The study of the properties of the Higgs boson is an urgent task, as evidenced by the experimental data of the ATLAS and CMS collaborations. We have presented the actual experimental data of the cross sections of Higgs boson decay into b-quarks  in associated production with a top-quark pair in  pp collisions at $\sqrt{s}$ = 13 TeV and an integrated luminosity of 139 fb$^{-1}$ with the ATLAS detector. This result corresponds to an observed (expected) significance of 1.0 (2.7) standard deviations. To clarify the ambiguities, we decided to consider the minimal extension of SM – MSSM model. We used the tree-level Higgs sector which can be described by two parameters M$_A$ and tan$\beta$. The calculation of BR($h\rightarrow b\bar{b}$) as the function of M$_A$ and production cross section of both Higgs bosons as the function of tan$\beta$ gives us the possibility to choose the optimal parameter space corresponding to the maximum values of BR and the production cross section. Using received parameters (M$_A$ = 200, tan$\beta$=2) we calculated cross sections of associated $t\bar{t}h(H)$ production at 13 and 14 TeV and the corresponding kinematical cuts on transverse momentum and rapidity. We found out the value of the mass of h (126 GeV) and H (330 GeV) at the chosen parameters from the constructed mass distribution. The values of BR($h\rightarrow b\bar{b}$) and BR($H\rightarrow b\bar{b}$)  are equal to 0.85 and 0.05 correspondingly.

\end{document}